# The Price Impact of Generalized Order Flow Imbalance

Yuhan Su[a,b], Zeyu Sun[a], Jiarong Li[b], Xianghui Yuan[a,*]

[a]School of Economics and Finance, Xi'an Jiaotong University, Xi'an 710000, China
[b]School of Software Engineering, Xi'an Jiaotong University, Xi'an 710000, China

**Abstract**
Order flow imbalance can explain short-term changes in stock price. This paper considers the change of non-minimum quotation units in real transactions, and proposes a generalized order flow imbalance construction method to improve Order Flow Imbalance (OFI) and Stationarized Order Flow Imbalance (log-OFI). Based on the high-frequency order book snapshot data, we conducted an empirical analysis of the CSI 500 constituent stocks. In order to facilitate the presentation, we selected 10 stocks for comparison. The two indicators after the improvement of the generalized order flow imbalance construction method both show a better ability to explain changes in stock prices. Especially Generalized Stationarized Order Flow Imbalance (log-GOFI), using a linear regression model, on the time scales of 30 seconds, 1 minute, and 5 minutes, the average $R^2$ out of sample compared with Order Flow Imbalance (OFI) 32.89%, 38.13% and 42.57%, respectively increased to 83.57%, 85.37% and 86.01%. In addition, we found that the interpretability of Generalized Stationarized Order Flow Imbalance (log-GOFI) showed stronger stability on all three time scales.

*Keywords*：High-frequency Trading, Limit Order Book, Chinese Financial Market, Order Flow Imbalance

## 1. Introduction

The maturity of China's electronic trading system and the openness of high-frequency data information have made high-frequency trading increasingly emerging in China's financial market. The availability of high-frequency trading and quotation records has inspired a large amount of empirical and theoretical literature on the relationship between order flow, liquidity, and price movements in order-driven market [1]. A particularly important issue in practical applications is the impact of orders on prices: within a specific time frame, the assumption of the best liquidation process that affects stock prices is proposed, which laid a solid foundation for the optimal execution problem with price impact [2-4]. Ma et al. [5] discussed the optimal portfolio execution problem under the influence of random prices. But with the rise of high-frequency trading, the order book based on high-frequency trading provides valuable information for exploring changes in stock prices. High-frequency trading is a major innovation in the financial market, accounting for a high proportion in stock trading in the United States and Europe [6]. O'Hara [7] pointed out that technology and high-frequency trading have changed the market, and widely discussed the impact of these changes on the microstructure of high-frequency market. Jones [8] suggested that high-frequency trading does not bring any instability to the market, and can even improve the overall quality of the market and reduce transaction costs. Brogaard et al. [9] examine the role of high-frequency traders (HFTs) in price discovery and price efficiency. Overall HFTs facilitate price efficiency. According to Berger et al. [10], the existence of the high-frequency trading program has significantly increased the trading volume and the depth of the bid price, and its impact on the market is benign. It can even be said to have a certain positive effect from the perspective of promoting trading volume and improving the depth of the bid price. Chinese stock exchanges provide snapshot data of the order book with a frequency of 3 seconds. The snapshot data of the limit order book discloses investors' information on the short-term market micro-state. The limit order book data fully demonstrates the competitive behavior of different investors. Therefore, a better understanding of how the structure of the limit order book affect the price formation has theoretical and practical significance.

Cao and Hansch et al. [11] proposed Depth Imbalance (QR) and Width Imbalance (HR) to describe the shape of the limit order book. Cont et al. [12] proposed Order Flow Imbalance, and found that the change of stock intermediate price can be explained by order book imbalance, and the interpretability is better than Trade Imbalance (VOL). The transaction Volume Order Flow Imbalance (VOI) constructed by Shen [13] is a feature that measures the incremental difference in the order volume at the optimal buying and selling price within a certain period of time, reflecting the supply and demand of investment behavior at the optimal buying and selling price. Xu et al. [14] proposed Multi-level Order Flow Imbalance (MLOFI), which is a vector that measures the net flow of buy and sell orders at different price levels in a limit order book, and further described the process of order-flow activity in the limit order book affect the price-formation. Sirignano and Cont [15] use path

---

* Corresponding author.

E-mail address: xhyuan@mail.xjtu.edu.cn

dependence to explain that there is a universal and stationary relationship between historical order flow data and the direction of price movements. Wang et al. [16] proposed Stationarized Order Flow Imbalance (log-OFI) based on Order Flow Imbalance, which improved the interpretability of the original indicator and realized a statistical arbitrage strategy based on this indicator.

Order flow imbalance is established in a given time interval. The optimal execution price of buyers and sellers is calculated strictly on the basis of the minimum quotation unit change, and the order information of the non-optimal execution price is not well used. In this paper, we propose a generalized construction method for order flow imbalance. In the second part, we describe the structure of Order Flow Imbalance and Stationarized Order Flow Imbalance, and propose that the optimal execution price between adjacent high-frequency order snapshots does not change according to the minimum quotation unit. We have carried out a generalized construction of the original order flow imbalance, and propose our generalized stationarized order flow imbalance. In the third part, we select 10 CSI 500 constituent stocks to conduct empirical analysis on four indicators: Order Flow Imbalance, Stationarized Order Flow Imbalance, Generalized Order Flow Imbalance, and Generalized Stationarized Order Flow Imbalance, evaluating their interpretability to the stock price changes.

## 2. Generalized order flow imbalance

### 2.1. Order flow imbalance and Stationarized order flow imbalance

The limit order book is composed of different timestamps, execution prices, and order quantities corresponding to the execution prices. It is an important tool for analyzing the behavior of market participants. Cont [1] focused in particular on models that describe the limit order book as a queuing system. First, simplify the working mode of the order book. For the order book, it is assumed that the number of orders at each price level does not exceed $D$ at most. When the order quantity corresponding to the optimal buyer's execution price or seller's execution price reaches $D$, the order book will create a new optimal buyer's execution price or seller's execution price. At this time, the number of orders placed, the number of orders cancelled, and the number of transactions will be accumulated along the new price until the quantity reach $D$ again or zero, so that the optimal buyer's execution price or the seller's execution price will move anew. Taking the entry of a new buyer order as an example, the image description is as follows:

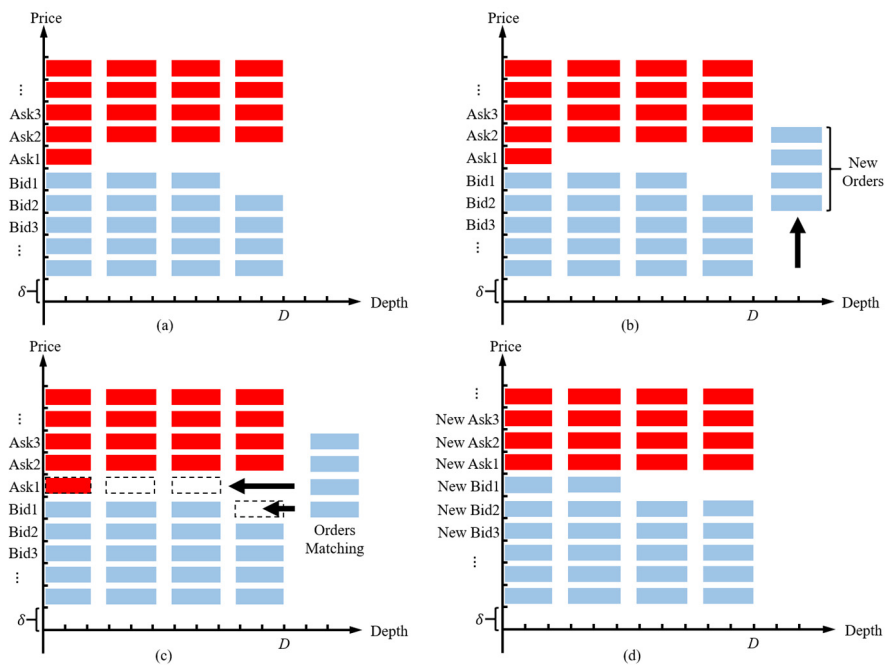

Figure 1: The buyer's new orders arrive and the price moves while maximum depth at each price level is equal to $D$.

As shown in the figure, the order book is in the state (a) at the initial moment, and when a new batch of buyer limit orders arrive (b), the number of orders on the optimal buyer's execution price will increase. If the number of orders on the optimal buyer's execution price exceeds the D, the orders arriving in the future will be accumulated along the new price, as shown in Figure (c). It should be noted that there are some seller orders at the new price level in the figure, so buyer orders arriving here will match the seller's orders according to the trading rules. In the end, the strength of the buyer and seller reaches a new balance, and a new order book (d) can be obtained at this time. This is just an example of the assumptions of the model.

Based on the above settings, Cont proposed a linear model describing order flow imbalance and price changes [12]. Consider a certain time interval $[t_{k-1}, t_k]$, during this period of time, the number of orders that reach the current optimal buyer's execution price is recorded as $L^b_{k-1,k}$, and the number of cancelled orders is recorded as $C^b_{k-1,k}$, and the number of transactions that match the market orders from the seller is recorded as $M^s_{k-1,k}$. The relationship between the change of the buyer's order and the change of the optimal buyer's execution price can be obtained:

$$\Delta P^b_{k-1,k} = \delta \left[ \frac{L^b_{k-1,k} - C^b_{k-1,k} - M^a_{k-1,k}}{D} \right] \quad (1)$$

Where δ represents the transaction price unit. In the same way, the relationship between the change in the seller's order and the change in the optimal seller's execution price can be obtained, which differs only in the direction of buying and selling:

$$\Delta P^a_{k-1,k} = -\delta \left[ \frac{L^a_{k-1,k} - C^a_{k-1,k} - M^b_{k-1,k}}{D} \right] \quad (2)$$

In the above formula, $\Delta P^b_{k-1,k} = P^b_k - P^b_{k-1}$, $\Delta P^a_{k-1,k} = P^a_k - P^a_{k-1}$. $P^b$ represents the optimal buyer's execution price, and $P^a$ represents the optimal seller's execution price. Define the intermediate price as $P_k = (P^b_k + P^a_k)/2\delta$, and define Order Flow Imbalance in this period of time as $OFI_{k-1,k} = L^b_{k-1,k} - C^b_{k-1,k} - M^a_{k-1,k} - L^a_{k-1,k} + C^a_{k-1,k} + M^b_{k-1,k}$, where $\varepsilon_k$ is used to represent the truncation error, we can get:

$$\Delta P_{k-1,k} = \frac{OFI_{k-1,k}}{2D} + \varepsilon_k \quad (3)$$

The model shows that there is a linear relationship between the change in the mid-price and Order Flow Imbalance. To test this relationship from an empirical point of view, Cont gave a measure of order flow imbalance. Still considering the time interval $[t_{k-1}, t_k]$, divide this period of time into N small observation intervals [12], and define:

$$OFI_{k-1,k} = \sum_{n=1}^{N} e_n \quad (4)$$

$$e_n = \Delta Bid_n - \Delta Ask_n \quad (5)$$

$$\Delta Bid_n = \begin{cases} q^b_n, & P^b_n > P^b_{n-1} \\ q^b_n - q^b_{n-1}, & P^b_n = P^b_{n-1} \\ -q^b_{n-1}, & P^b_n < P^b_{n-1} \end{cases} \quad \Delta Ask_n = \begin{cases} -q^a_{n-1}, & P^a_n > P^a_{n-1} \\ q^a_n - q^a_{n-1}, & P^a_n = P^a_{n-1} \\ q^a_n, & P^a_n < P^a_{n-1} \end{cases} \quad (6)$$

Among them, $q^b_{n-1}$ and $q^b_n$ represent the optimal buyer's execution price order quantity at time $n-1$ and time $n$, respectively, and $q^a_{n-1}$ and $q^a_n$ represent the optimal seller's execution price at time $n-1$ and time $n$. $\Delta Bid_n$ represents the increase in buyer power in the nth observation interval, and $\Delta Ask_n$ represents the increase in seller power in the nth observation interval.

Note that the linear model imposes a limit on the number of orders at each price level, but in fact this limit does not exist. The number of orders corresponding to different levels is very different, showing high volatility, which to some extent will cause the model to be inaccurate. Wang et al. [16] have performed logarithmic processing on the classic order flow imbalance indicator to make the data show higher stability. This scheme can effectively reduce the bias caused by the threshold hypothesis to the empirical analysis, which means that the maximum depth at each price level is equal to D. Based on the symbols mentioned above, the stationarized order flow imbalance (log-OFI) is to take the logarithm of each q on the basis of the original OFI.

**2.2. Construct a generalized order flow imbalance**

Considering that there is a minimum observation interval in the real market (for example, the time interval of order book snapshot data provided by the Chinese stock market is three seconds), the size of N used to divide the observation interval cannot be chosen arbitrarily, which will largely lead to discontinuous movement of the optimal execution price (movement exceeds one $\delta$). Discuss the process of increasing the buyer's optimal execution price as exemplified before. Since the entire process usually occurs in a short time, for a three-second observation interval, such a process may have occurred multiple times. This will cause the optimal buyer's execution price to move beyond one $\delta$.

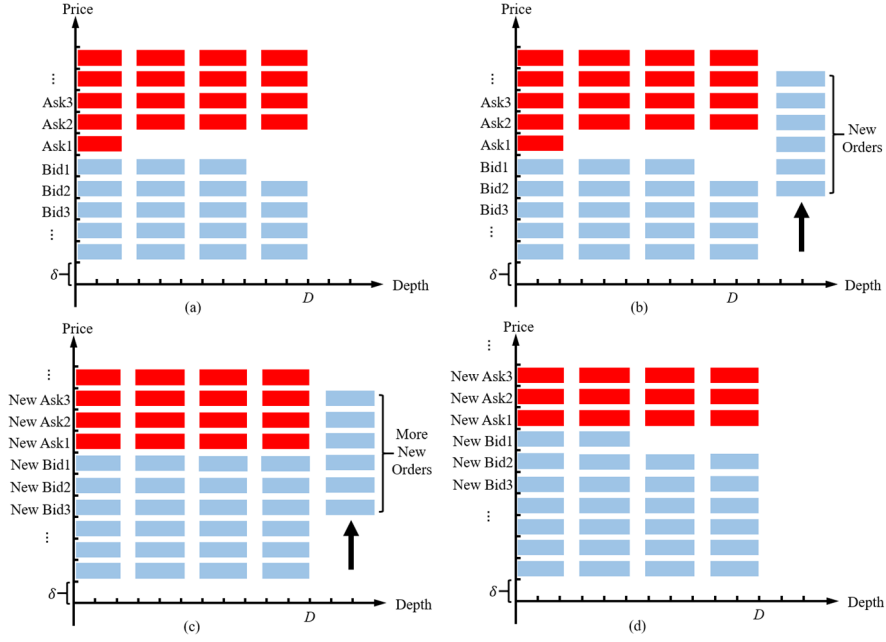

Figure 2: Buyer's new order arrival and optimal execution price movement in a small observation interval while maximum depth at each price level is equal to *D*.

The figure ignores the previously described order matching step, and directly gives the corresponding results, showing the situation where the buyer's optimal execution price moves by 2δ, all of which happen in a short time. In the figure (a) and (d) are the initial and final conditions that can be observed in the interval, and (b) and (c) indicate that two batches of new orders have entered during this period. It should be noted that the image is only an illustration, and there are many situations that cause the optimal quotation to move similar to this.

In response to this situation, this paper relaxes the restriction that the optimal price within the observation interval implied by the order flow imbalance can only move one δ, and proposes a new order flow indicator called Generalized Order Flow Imbalance (GOFI). This indicator is no longer based on the position of the optimal execution price, but focuses on the value of the optimal execution price. Consider the time interval $[t_{k-1}, t_k]$, divide this period into N small observation intervals, Generalized Order Flow Imbalance is defined as follows:

$$GOFI_{k-1,k} = \sum_{n=1}^{N} gene_n \qquad (7)$$

$$gene_n = \Delta GBid_n - \Delta GAsk_n \qquad (8)$$

$$\Delta GBid_n = \begin{cases} \sum_{m^b} v_{n,i}^b - v_{n-1,1}^b, & P_n^b > P_{n-1}^b \\ v_{n,1}^b - v_{n-1,1}^b, & P_n^b = P_{n-1}^b \\ v_{n,1}^b - \sum_{m^b} v_{n-1,i}^b, & P_n^b < P_{n-1}^b \end{cases} \quad \Delta GAsk_n = \begin{cases} v_{n,1}^a - \sum_{m^a} v_{n-1,i}^a, & P_n^a > P_{n-1}^a \\ v_{n,1}^a - v_{n-1,1}^a, & P_n^a = P_{n-1}^a \\ \sum_{m^a} v_{n,i}^a - v_{n-1,1}^a, & P_n^a < P_{n-1}^a \end{cases} \qquad (9)$$

In the formula, $\Delta GBid_n$ reflects the increase in buyer power in a small observation interval, $\Delta GAsk_n$ reflects the increase in seller power in a small observation interval, and the two can be subtracted to measure generalized order flow imbalance within a small observation interval. To explain the newly appeared symbols in the above formula, still taking the buyer as an example, $v_{n,i}^b$ means that at the n time of the nth small observation interval, the quantity of buyer orders at the ith price level. $v_{n-1,i}^b$ means that at the n-1 time of the nth small observation interval, the quantity of buyer orders at the ith price level. In the same way, the meanings of $v_{n,i}^a$ and $v_{n-1,i}^a$ can be obtained. There are two cases involved here. When $v$ uses orders quantity directly, the indicator obtained is $GOFI_{k-1,k}$; considering the stationarity correction, when $v$ takes the logarithmic value of the orders quantity, the indicator obtained is log-$GOFI_{k-1,k}$. For example, if $i = 1$, it means that $v$ represents the (logarithmic) value of the orders quantity at the buyer's execution price or the seller's execution price; if $i = 2$, it means that

$v$ represents the (logarithmic) value of the orders quantity at the second bid price or the second ask price. $m$ is used to determine the value range of $i$ in the above formula, and its superscript expresses the buying and selling direction.

$$m^b = \left|\frac{P_n^b - P_{n-1}^b}{\delta}\right| + 1, \quad m^a = \left|\frac{P_n^a - P_{n-1}^a}{\delta}\right| + 1 \tag{10}$$

$m^b$ represents the number of all the stalls where the pending order price at time $n$ is greater than or equal to or less than or equal to the optimal buyer's execution price at time $n-1$ when the optimal buyer's execution price changes in the nth small observation interval. In short, $m$ measures the level of the optimal bid movement.

3. **Comparison of Interpretability Between Four Order Flow Imbalances and Mid-price Changes**

This paper uses the coefficient of determination obtained by linear regression to measure the strength of the linear relationship between the order flow imbalance and changes in the mid-price. In order to facilitate the presentation, we just select 10 stocks that are actively traded in the CSI 500 as the objects of empirical analysis. The in-sample time range is from January 1, 2021 to March 31, 2021, and the out-of-sample time range is from April 1, 2021 to June 31, 2021. The data only includes the continuous bidding period from 9:30:00 to 14:57:00, excluding the date of the price limit. Next, consider four indicators marked OFI, GOFI, log-OFI and log-GOFI, the time interval $[t_{k-1}, t_k]$ used to construct the indicator takes 30 seconds, 1 minute, and 5 minutes respectively. The regression equations corresponding to the tables and figures are as follows:

$$\Delta P_{k-1,k} = \beta OFI_{k,k-1} + \varepsilon_k \tag{11}$$

$$\Delta P_{k-1,k} = \beta GOFI_{k,k-1} + \varepsilon_k \tag{12}$$

$$\Delta P_{k-1,k} = \beta log\text{-}OFI_{k,k-1} + \varepsilon_k \tag{13}$$

$$\Delta P_{k-1,k} = \beta log\text{-}GOFI_{k,k-1} + \varepsilon_k \tag{14}$$

Table 1-3 shows the coefficient of determination of the four order flow imbalances in the sample and out of sample.

Table 1: The coefficient of determination of the linear regression between four order flow imbalances and mid-price change within 30 seconds.

| StockCode | $OFI(R_{IS}^2)$ | $OFI(R_{OOS}^2)$ | $GOFI(R_{IS}^2)$ | $GOFI(R_{OOS}^2)$ | $log\text{-}OFI(R_{IS}^2)$ | $log\text{-}OFI(R_{OOS}^2)$ | $log\text{-}GOFI(R_{IS}^2)$ | $log\text{-}GOFI(R_{OOS}^2)$ |
|---|---|---|---|---|---|---|---|---|
| 000998.XSHE | 39.65 | 10.69 | 49.19 | 24.34 | 75.83 | 78.59 | 86.83 | 90.36 |
| 002382.XSHE | 39.52 | 32.6 | 51.31 | 43.56 | 72.86 | 75.90 | 84.91 | 88.13 |
| 300088.XSHE | 38.06 | 28.94 | 41.60 | 33.6 | 93.20 | 93.22 | 96.82 | 97.02 |
| 300146.XSHE | 35.27 | 30.38 | 44.03 | 38.63 | 74.89 | 70.24 | 85.27 | 80.09 |
| 300223.XSHE | 32.06 | 36.12 | 38.94 | 40.79 | 63.05 | 59.44 | 65.99 | 60.46 |
| 600089.XSHG | 42.41 | 36.59 | 51.37 | 47.01 | 80.30 | 85.71 | 91.37 | 95.62 |
| 600166.XSHG | 40.61 | 37.38 | 43.40 | 40.66 | 93.01 | 94.09 | 97.58 | 97.42 |
| 600460.XSHG | 43.67 | 40.44 | 51.77 | 49.09 | 72.02 | 69.54 | 80.67 | 77.30 |
| 600988.XSHG | 38.15 | 39.85 | 45.14 | 45.35 | 76.37 | 79.00 | 85.83 | 89.02 |
| 603290.XSHG | 34.36 | 35.89 | 40.59 | 40.48 | 58.10 | 57.93 | 60.57 | 60.24 |
| Average | 38.38 | 32.89 | 45.73 | 40.35 | 75.96 | 76.37 | 83.58 | 83.57 |

Table 2: The coefficient of determination of the linear regression between four order flow imbalances and mid-price change within 1 minute.

| StockCode | $OFI(R^2_{IS})$ | $OFI(R^2_{OOS})$ | $GOFI(R^2_{IS})$ | $GOFI(R^2_{OOS})$ | $log\text{-}OFI(R^2_{IS})$ | $log\text{-}OFI(R^2_{OOS})$ | $log\text{-}GOFI(R^2_{IS})$ | $log\text{-}GOFI(R^2_{OOS})$ |
|---|---|---|---|---|---|---|---|---|
| 000998.XSHE | 46.63 | 12.18 | 55.63 | 29.34 | 78.72 | 79.38 | 89.04 | 91.58 |
| 002382.XSHE | 45.15 | 38.72 | 56.73 | 51.39 | 74.68 | 79.19 | 86.51 | 90.94 |
| 300088.XSHE | 48.31 | 37.72 | 52.31 | 42.37 | 93.78 | 93.55 | 97.60 | 97.65 |
| 300146.XSHE | 38.27 | 32.97 | 47.39 | 41.58 | 74.89 | 71.09 | 86.05 | 81.39 |
| 300223.XSHE | 37.49 | 41.69 | 44.54 | 46.86 | 64.18 | 61.48 | 68.30 | 63.79 |
| 600089.XSHG | 47.01 | 40.94 | 56.64 | 52.82 | 82.10 | 85.90 | 92.64 | 96.12 |
| 600166.XSHG | 47.24 | 45.87 | 50.46 | 49.12 | 92.88 | 94.64 | 98.09 | 98.09 |
| 600460.XSHG | 46.86 | 43.48 | 55.12 | 51.66 | 73.82 | 71.48 | 83.14 | 79.48 |
| 600988.XSHG | 43.97 | 45.86 | 50.48 | 51.51 | 78.02 | 80.88 | 87.15 | 90.90 |
| 603290.XSHG | 43.57 | 41.91 | 49.51 | 46.01 | 63.24 | 60.94 | 66.08 | 63.77 |
| Average | 44.45 | 38.13 | 51.88 | 46.27 | 77.63 | 77.85 | 85.46 | 85.37 |

Table 3: The coefficient of determination of the linear regression between four order flow imbalances and mid-price change within 5 minutes.

| StockCode | $OFI(R^2_{IS})$ | $OFI(R^2_{OOS})$ | $GOFI(R^2_{IS})$ | $GOFI(R^2_{OOS})$ | $log\text{-}OFI(R^2_{IS})$ | $log\text{-}OFI(R^2_{OOS})$ | $log\text{-}GOFI(R^2_{IS})$ | $log\text{-}GOFI(R^2_{OOS})$ |
|---|---|---|---|---|---|---|---|---|
| 000998.XSHE | 51.90 | 5.08 | 60.14 | 31.67 | 79.50 | 77.24 | 90.60 | 91.78 |
| 002382.XSHE | 50.41 | 43.15 | 61.22 | 57.17 | 75.49 | 81.58 | 87.04 | 92.42 |
| 300088.XSHE | 60.8 | 48.78 | 65.62 | 53.62 | 94.58 | 94.22 | 98.61 | 98.58 |
| 300146.XSHE | 42.10 | 34.06 | 51.58 | 43.36 | 71.50 | 67.24 | 85.49 | 81.07 |
| 300223.XSHE | 41.88 | 48.53 | 49.16 | 54.09 | 61.78 | 62.44 | 68.10 | 66.70 |
| 600089.XSHG | 49.08 | 46.24 | 59.36 | 57.78 | 81.63 | 85.96 | 93.08 | 96.25 |
| 600166.XSHG | 56.54 | 60.17 | 59.49 | 63.08 | 92.89 | 94.13 | 98.68 | 98.90 |
| 600460.XSHG | 44.60 | 44.62 | 53.96 | 51.79 | 70.27 | 70.14 | 81.58 | 79.06 |
| 600988.XSHG | 49.18 | 48.77 | 52.48 | 54.53 | 76.24 | 79.28 | 85.18 | 91.28 |
| 603290.XSHG | 49.92 | 46.27 | 54.61 | 49.45 | 66.47 | 61.37 | 69.22 | 64.03 |
| Average | 49.64 | 42.57 | 56.76 | 51.65 | 77.04 | 77.36 | 85.76 | 86.01 |

In order to intuitively compare the interpretability of four order flow imbalances on price changes, Figure 3-5 shows the coefficient of determination of 10 stocks. The four order flow imbalances have different levels of $R^2$: the regression equation based on log-GOFI generally has the best $R^2$, the next one is log-OFI, then GOFI and OFI. And the $R^2$ of log-GOFI will not change due to changes in the time scale, and has been maintained at a high level.

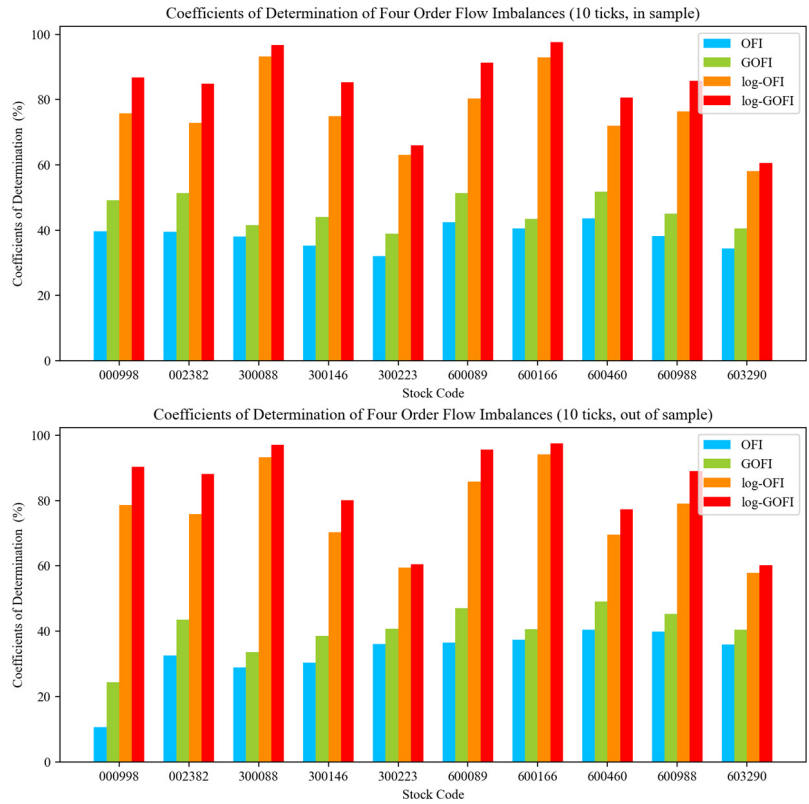

Figure 3: Comparison of the coefficient of determination of the linear regression between four order flow imbalances and mid-price change within 30 seconds.

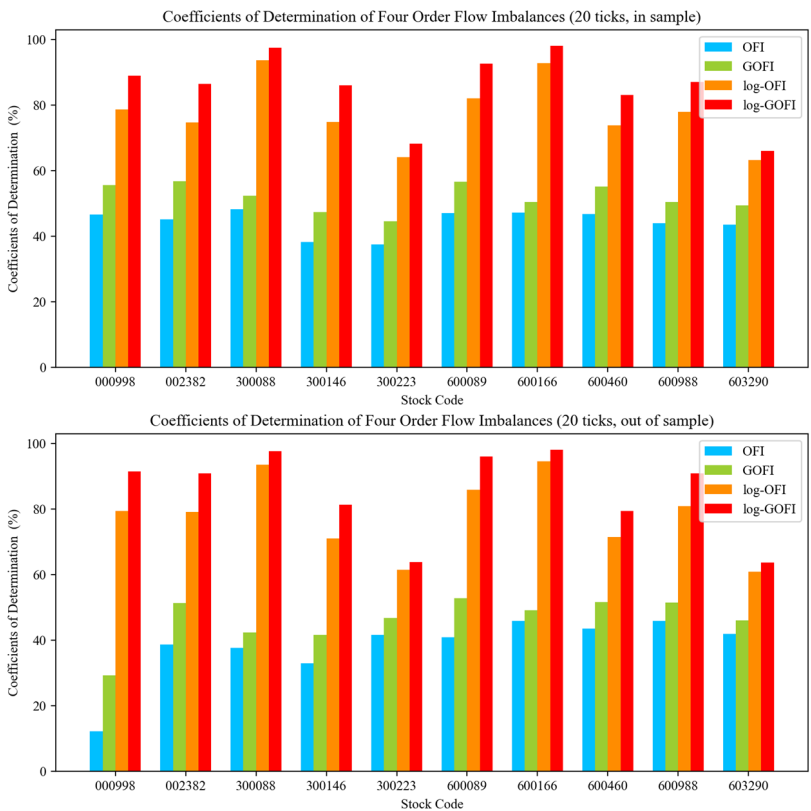

Figure 4: Comparison of the coefficient of determination of the linear regression between four order flow imbalances and mid-price change within 1 minute.

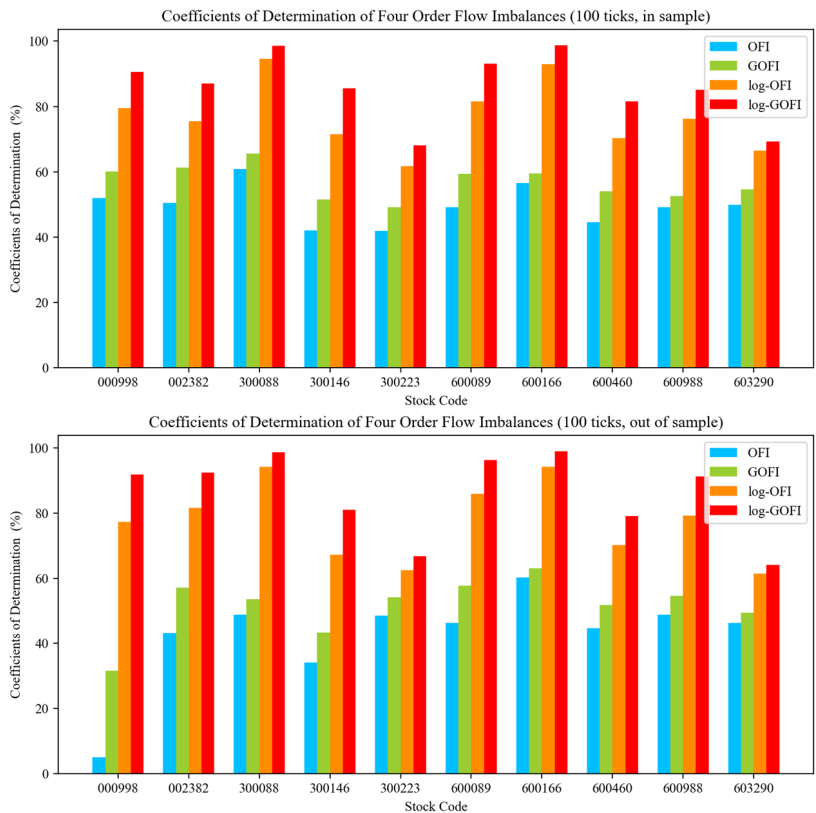

Figure 5: Comparison of the coefficient of determination of the linear regression between four order flow imbalances and mid-price change within 5 minutes.

The empirical results of this paper show that the generalized order flow imbalance has better explanatory performance than the classical order flow imbalance, and this superiority also manifests itself after the logarithmization of the indicators. From the perspective of the size of the time interval $[t_{k-1}, t_k]$, the $R^2$ average of all stock samples shows that with the increase in the length of construction time (from 30 seconds to 5 minutes), the in-sample interpretation performance of the four indicators has generally improved, and the same conclusion is shown out of the sample. From the comparison result of $R^2$ in-sample and out-of-sample, it can be found that the out-of-sample fitting abilities of the OFI and GOFI have decreased, but the out-of-sample fitting abilities of the log-OFI and log-GOFI remain stable. log-GOFI is the outstanding indicator among the four indicators. Take the 5-minutes indicator as an example, the in-sample average $R^2$ of the log-GOFI regression equation is 85.76%, and the out-of-sample average $R^2$ of the log-GOFI regression equation is 86.01%.

## 4. Conclusion

The simplified model of the order book links the order flow imbalance with changes in stock price, and provides ideas for high-frequency trading researchers to gain a deeper understanding of the microstructure of the stock market. This paper considers the change of non-minimum quotation unit in real trading, and proposes a generalized construction method of order flow imbalance. We use the generalized construction method to propose Generalized Order Flow Imbalance (GOFI) and Generalized Stationarized Order Flow Imbalance (log-GOFI). On this basis, this paper compares the performance of different order flow imbalances under different time scales, and compares the in-sample and out-of-sample performance of different order flow imbalances. The empirical analysis of the sample stocks shows that, compared with the original two indicators, the two new indicators that take into account the changes in the non-minimum quotation unit in real trading can obtain a higher linear regression $R^2$. And Generalized Stationarized Order Flow Imbalance (log-GOFI) has achieved a higher and more stable $R^2$ on different time scales or inside and outside the sample.